\documentstyle[12pt,epsf]{article}
\begin{document}
\sloppy
\title{The Pion Electroproduction in the $\Delta$(1232) Region}

\author{Chung Wen Kao\\
Department of Physics,
University of Maryland\\
College Park, MD 20742-4111\\
\footnotesize DOE/ER/40762-199~~~~UMD PP\#00-036}
\maketitle

\begin{abstract}
 The amplitudes of the pion electroproduction
  from the nucleon are derived up to ${\cal O}(\epsilon^3)$
 in the framework
of chiral effective theory including pions, nucleons
 and $\Delta$(1232) isobars as explicit degrees of freedom.
 The $Q^2$ evolutions of the
weighted integrals of amplitudes are presented and
the predictive power of the method proposed in the previous study
of pion photoproduction \cite{KC} is shown.  
\end{abstract}

\vspace{20. mm}

\pagebreak
\noindent
\section {Introduction}

With the advent of the new generation
of high intensity, high duty-factor electron accelerators as
MAMI(Mainz), ELSA(Bonn) and Jefferson Lab, as well as modern laser
backscattering facilities at LEGS(Brookhaven) and GRAAL(Grenoble),
a great amount of precise data probing the structure
of the nucleon at low energy has become available or
is expected in the near future.
Most of these data are from scattering processes with either the real
or virtual photons or with pions. 
 These different processes should not be treated separately
since they are intimately related.
 For example the Fermi-Watson theorem \cite{Wa}
relates the phase shift between the pion-nucleon scattering and
the pion photo- and electroproduction. In order
to understand the properties of the nucleon 
through recent data, a consistent framework
to describe these physical processes is essential.\\

In the low energy regime chiral symmetry largely governs the dynamics
of the pions and nucleons by severely restricting the interactions
between them; chiral symmetry ought
to play a crucial role
in the analysis of various experimental data of the $\gamma N\pi$ system.
It turns out that Heavy Baryon Chiral Perturbation Theory (HBChPT) becomes
a very efficient and powerful tool to study the behavior of $N\pi$ system
in the low energy region because
HBChPT automatically  satisfies the constraint
of chiral symmetry at the leading order and provides a systematic
way of including the effects of finite quark mass and other sectors
at higher energies. Many calculations of HBChPT on the threshold
were done and most were proven successful \cite{BKM}.\\

However the validity of the original HBChPT
 beyond the threshold is threatened
by the existence of the nucleon excited states, ${\it i.e.}$, the resonances.
 In the original HBChPT such states are
 integrated out and their effects are replaced by
 the finite piece of counterterms,
therefore the direct connection is lost.
Phenomenologically this is a reasonable scheme
so long as these resonances are heavy compared to the energy
scale of interest. 
However for the case of $\Delta$(1232) resonance it is questionable
because $\Delta$(1232) is light and strongly couples to the $N\pi$ system.
This is consistent with Large $N_{c}$ QCD which requires a light,
 strongly coupled I=J=3/2 state
\cite{JM,JM2}. 
Therefore it is sensible to include the $\Delta$(1232) as an explicit
degree of freedom in the effective chiral Lagrangian
 applied in the $\Delta$(1232) region.
 This phenomenological
extension of HBChPT is presumably a reasonable and
consistent  scheme to investigate the
processes such as $\gamma N\rightarrow \pi N$, $\pi N\rightarrow \pi N$
and $\gamma N \rightarrow \gamma N$ in the $\Delta$(1232) region
 \cite{BSS,BSS2,HHK,HHK2}, and the relations between these processes
are reflected by the fact that they share many of the counterterms.\\

   Resonances themselves are the subjects
 of intense interest and study. 
There are tremendous amount of activities, both theoretical and
experimental, trying to extract the information
 about the $\Delta(1232) \rightarrow
N$ electromagnetic transition;  
 it is relevant for models that made specific statements about
the internal wave functions of baryons. For instance, the most
 naive constituent
models predicted the ratio
of E2/M1 and C2/M1 to be both zero because the nucleon and $\Delta$
isobars are perfectly spherical since the spatial parts of their
wavefunctions are both ground states. 
The one-gluon-exchange tensor
forces only gives very small values of E2/M1 and C2/M1,
and different baryon models give different values generated by different
mechanisms.
 Recently 
 the new $p(\vec{e},e^{'}\vec{p})\pi^{0}$
 experiments at MAMI and MIT/Bates, and those planned at CEBAF, have 
raised the interest in the theoretical calculations of the amplitudes for
electroproduction of low lying baryon resonances and it was expected
that the new experimental effort would improve our understanding
on the N$\rightarrow \Delta$ transition \cite{M,S,S2,W,Ka,ST}.\\

 However, 
the new data strongly
suggests that nonresonant amplitudes contribute significantly
both in the longitudinal and transverse channels \cite{M}.
Then the interpretation of the experimentally determined $R_{EM}$ and
$R_{SM}$ is severely complicated by the presence of processes which
are coherent with the $\Delta$(1232) resonance. These processes
give rise to additional quadrupole amplitudes in the invariant mass region
of interest and contaminate previous $R_{EM}$ and $R_{SM}$, so that
the extraction of information about $\Delta$(1232)
turns out to be more difficult than expected.
It was also
demonstrated that all available calculations exhibit deviation
from the data. Thus an improved theoretical framework which can describe
 both resonant and nonresonant amplitudes simultaneously is
mandated. The approach based on the effective chiral Lagrangian
including $\Delta$(1232) is a promising choice because we made no
assumption on the magnitudes of the background contributions
and the $\Delta$(1232) isobars are treated
 explicitly with the nucleons. 
\\

 Actually, it has already been shown that
there is an intrinsic theoretical difficulty to separate
the resonant contribution with the
background part due to the ambiguity of unitarity \cite{WWA}.
 The approach based on HBChPT
certainly is free of this kind of ambiguity
since HBChPT is unitary order by order. 
But it turns out another ambiguity emerges:
 the $\gamma$N$\Delta$ coupling constants
 have to absorb the divergences
generated from the loop diagrams and be renormalized
infinitely.
 Although final amplitudes do not depend on the
 renormalization scheme, the separation between the resonant and
 the background
part does. Since there is no on-shell $\Delta$ isobar
 available in the lab,  
this kind of separation becomes arbitrary. Therefore any statement about
$\Delta$(1232) resonance in our framework becomes scheme dependent.
Therefore instead of 
the $N \rightarrow \Delta$ transition we calculated the amplitudes of
electroproduction of pions in the $\Delta$(1232) 
region which in principle can be measured
 without any ambiguity.\\

 Unfortunately the procedure of extraction of the $\gamma$N$\Delta$ 
coupling constants would not
be straightforward.
 In our previous study on the pion photoproduction, we pointed
out that the results of HBChPT could not directly compare with experimental 
data because in the $\Delta$(1232) region the 
straightforward power counting scheme
breaks down as some amplitudes become uncontrollable 
around the $\Delta$(1232) pole. One natural way to cure it is to
put the self energy of $\Delta$(1232) in the propogator, then the $\Delta$
pole is removed from the real axis and the $\delta$ function in the imaginary
part of the propagator is also smoothed.
 However this manipulation makes
 power counting unreliable, if not impossible. Such a manipulation
is not allowed
if the formal structure of the power
 counting scheme is required. This appears to be true
in our case because the smallness of $R_{EM}$ and $R_{SM}$ implies that the
$\gamma$N$\Delta$ vertices we are studying is very weak, 
and therefore the error of theoretical
 calculation is crucial for the verification. Thus one
reasonable way to compare the HBChPT results with experimental data is via
weighted integrals of the amplitudes through the $\Delta$(1232) region.
So we cannot make predictions on the physical observables directly,
but only on the integrated quantities about
 the amplitudes of the physical processes. It becomes the main limitation
of our approach.\\

 The same limitation
remains in the case of electroproduction of pions. The $\Delta$(1232) pole
 also exists in electroproduction amplitudes except in S-wave.
 Again we cannot
explicitly write down any $Q^{2}$ dependences of the physical observables.
 Unlike some model calculation,
 our approach cannot be generalized to the high $Q^2$
region without any modifications.
 But it was found that  
in HBChPT the photo- and electroproduction of pions
 are both determined by almost the same set of counterterms. In other
words, once the counterterms are fit by the data of photoproduction,
then the $Q^2$ evolution of the amplitudes of 
electroproduction is almost fixed in HBChPT,
and the $Q^2$ dependences of weighted integrals of
 those multipoles still provide a very
good testing ground for this chiral effective theory. \\

This article is organized as following: In Sec. 2 the general formalism
is briefly sketched. Section 3 discusses the renormalization. The results
of weighted integrals are 
given in Sec 4. We summarize and provide our perspective for further
efforts in Sec 5.\\

\section {Formulation}

Heavy Baryon Chiral Perturbation Theory (HBChPT) has been quite
 successful for scattering processes off a single nucleon
 near threshold \cite{BKM}.
The extended formulation including $\Delta$(1232) isobar as explicit degrees
of freedom is also well developed.
The basic idea is to treat $\Delta$=$m_{\Delta}-m_{N}$ as a light
scale as $m_{\pi}$ and expand the $S$ matrix by
 $\epsilon$=$\{\frac{m_{\pi}}{\Lambda},\,
\frac{p}{\Lambda},\,\frac{\Delta}{\Lambda}\}$, where $\Lambda$ is
the heavy scale typically of order such as  $4\pi{\it f}_{\pi}$
or $M_{p}$.
For details we refer the readers to extensive literature of reviews.
\cite{HHK, HHK2,GHKP}\\

\noindent
The procedure for the calculation of pion electroproduction 
in this chiral effective theory is similar
to the one of pion photoproduction. The only new
 $\gamma N\Delta$ vertex is:

$$\frac{-G_{3}}{(2m_{N})^{2}}\bar{\psi}^{\mu}_{i}
\Theta_{\mu\nu}(y_{1})\gamma_{5}Tr(\tau^{i}
D_{\rho}f_{+}^{\rho\nu})\psi_N+h.c.$$

Which vanishes in the real photon case 
but contributes in the virtual photon case. 
However it turns out that its effect starts from ${\cal O}(\epsilon^{4})$ amplitudes, therefore
all the Feymann rules we need are already in our previous work \cite{KC} and
the whole calculation is straightforward. \\

To extract
 the multipole results, the amplitudes are decomposed
 into the standard
CGLN amplitudes in the $\pi$N c.m. frame \cite{CGLN}:
\begin{equation}
\begin{array}[c]{c}
{\cal T}=i(\vec{\sigma}\cdot \vec{\epsilon}){\cal T}_{1}+
(\vec{\sigma}\cdot\hat{q})(\vec{\sigma}\cdot\hat{k}\times\vec{\epsilon})
{\cal T}_{2}+i(\vec{\sigma}\cdot \hat{k})(\hat{q}\cdot \vec{\epsilon})
{\cal T}_{3}+i(\vec{\sigma}\cdot \hat{q})(\vec{\epsilon} \cdot \hat{q})
{\cal T}_{4}\\
+i(\vec{\sigma}\cdot \hat{k})(\hat{k}\cdot \vec{\epsilon})
{\cal T}_{5}+i(\vec{\sigma}\cdot \hat{q})(\hat{k}\cdot\vec{\epsilon})
{\cal T}_{6}
-i(\vec{\sigma}\cdot \hat{q})\epsilon_{0}{\cal T}_{7}
-i(\vec{\sigma}\cdot \hat{k})\epsilon_{0}{\cal T}_{8}.
\end{array}
\end{equation}

\noindent
Note ${\cal T}_{5}$, ${\cal T}_{6}$, ${\cal T}_{7}$ and ${\cal T}_{8}$
vanish in pion photoproduction.
Due to the current conservation there are two relations:
$$|{\bf \vec{k}}|{\cal T}_{5}=k_{0}{\cal T}_{8};\,\,\,\,|{\bf \vec{k}}|{\cal T}_{6}=k_{0}{\cal T}_{7}.$$
So only six amplitudes are independent. Since our gauge condition is 
$\epsilon \cdot v$=0 and we choose $v_{\mu}$=(1,$\vec{0}$),  
the amplitudes are simplified as:

\begin{equation}
\begin{array}[c]{c}
{\cal T}=i(\vec{\sigma}\cdot \vec{\epsilon}){\cal T}_{1}+
(\vec{\sigma}\cdot\hat{q})(\vec{\sigma}\cdot\hat{k}\times\vec{\epsilon})
{\cal T}_{2}+i(\vec{\sigma}\cdot \hat{k})(\hat{q}\cdot \vec{\epsilon})
{\cal T}_{3}
+i(\vec{\sigma}\cdot \hat{q})(\vec{\epsilon} \cdot \hat{q})
{\cal T}_{4}\\
+i(\vec{\sigma}\cdot \hat{k})(\hat{k}\cdot \vec{\epsilon})
{\cal T}_{5}+i(\vec{\sigma}\cdot \hat{q})(\hat{k}\cdot\vec{\epsilon})
{\cal T}_{6}.
\end{array}
\end{equation}

The amplitudes are usually expressed in terms of three types of multipoles
: electric($E_{l\pm}$), magnetic$(M_{l\pm}$) and
 longitudinal ($L_{l\pm}$), 
with pion angular momentum $l$ and total momentum $j=l\pm 1/2$.
They can be calculated by inverting the following relations:
\begin{eqnarray}
{\cal T}_{1}&=&\sum_{l\geq 0}[ (lM_{l+}+E_{l+})P^{'}_{l+1}+
[(l+1)M_{l-}+E_{l-}]P^{'}_{l-1}],\\
{\cal T}_{2}&=&\sum_{l\geq 1} [(l+1)M_{l+}+lM_{l-}]P_{l}^{'},\\
{\cal T}_{3}&=&\sum_{l\geq 1}[(E_{l+}-M_{l+}) P^{''}_{l+1}+(E_{l-}+M_{l-})P^{''}_{l-1}],\\
{\cal T}_{4}&=&\sum_{l\geq 2} (M_{l+}-E_{l+}-M_{l-}-E_{l-})P^{''}_{l},\\
{\cal T}_{5}&=&\sum_{l\geq 0} [(l+1)L_{l+}P^{'}_{l+1}-lL_{l-}P^{'}_{l-1}],\\
{\cal T}_{6}&=&\sum_{l\geq 1} [(lL_{l-}-(l+1)L_{l+})P^{'}_{l}].
\end{eqnarray}

\noindent
Here $P^{'}_{l}$ are derivates of Legendre polynomials.
Note that in the literature the longitudinal transitions are often
described by $S_{l\pm}$ scalar multipoles which correspond to the multipole
decomposition of the amplitudes ${\cal T}_{7}$, ${\cal T}_{8}$. They are
connected with the longitudinal ones by $S_{l\pm}=|{\bf k}|L_{l\pm}/k_{0}$.\\

All of the observables are products of these amplitudes.
In general there are 16 different polarization observables for the
reaction with real photons.\cite{ALT}
 For the virtual photon we have four
additional ones due to longitudinal amplitudes and 16 observables due to
longitudinal-transverse interference. Thus there are 36 observables for pion
electroproduction. In view of the great number of possible polarization
 observables it is natural to ask which set of observables can be, in principle
,a complete determination of all amplitudes. Naively it may be argued that
any set of 11 observables should suffice to determine all amplitudes because
 there are six independent complex variables. But one overall phase is
 undetermined. However all observables are the products of amplitudes
 therefore the discrete ambiguity prevents us to take an arbitrary ones but 
 properly chosen set of observables to satisfy some criterion \cite{ALT}. \\

\section {Renormalization}
Our calculation contains several N$\pi$ and $\Delta$$\pi$ loop diagrams
which are regularized by dimensional regularization.
Their divergences 
must be absorbed by the counterterms
 $b_{i}$ of ${\cal L}_{N}^{(3)}$ or $G_{i}$ of ${\cal L}_{N\Delta}^{(3)}$.
The renormalization of the amplitudes of pion photoproduction
was discussed in detail in \cite{KC}. 
In the neutral pion electroproduction no new counterterms are needed.
In the charged pion electroproduction, two new parameters
$b_{7}$, and $b_{23}$ emerges. They are the coefficients of the
following counterterms in ${\cal L}_{\pi N}^{(3)}$ \cite{EM} respectively:
$${\cal O}_{7}=[D^{\mu},f_{+\mu\nu}]v^{\nu},\,\,{\cal O}_{23}=
S^{\mu}[D^{\nu},f_{-\mu\nu}].$$ Furthermore it was found that
\begin{equation}
\beta_{7}=\frac{1}{6}+\frac{5}{6}g_{A}^{2}-\frac{20}{27}g_{\pi\Delta N}^{2}.
\end{equation}
Here
$$b_{i}=b^{r}_{i}(\mu)+(4\pi)^{2}\beta_{i}L $$
$$L\equiv\frac{\mu^{d-4}}{(4\pi)^{2}}(\frac{1}{d-4}-\frac{1}{2}[\ln(4\pi)+1+\Gamma^{'}(1)]).$$
$b_{7}^{r}$ is related to the electric mean square charge radii of the proton. 
 Consider the nucleon matrix element
 of the isovector component of the quark vector current:
$$\langle N( p^{'})|\bar{q}\gamma_{\mu}\frac{\tau^{a}}{2}q|N(p)\rangle =
\bar{u}({p^{'}})[\gamma_{\mu}F_{1}^{V}(q^{2})+\frac{i\sigma_{\mu\nu}
q^{\nu}}{2m_{p}}F^{V}_{2}(q^{2})]\frac{\tau^{a}}{2}u(p),$$
with $q=p^{'}-p$. 
$$\langle r^{2}\rangle =6\frac{dF_{1}^{V}(q^2)}{dq^2}|_{q^{2}=0}.$$
The relation was given by \cite{BFHM}:
\begin{equation}
\begin{array}[c]{c}
\langle r^{2}\rangle_{1}=-\frac{1}{(4\pi F_{\pi})^{2}}\{1+7g_{A}^{2}+
(10 g_{A}^{2}+2)\ln(\frac{m_{\pi}}{\mu})\}
-\frac{12 b_{7}^{r}(\mu)}{(4\pi F_{\pi})^{2}}\\
+\frac{g_{\pi\Delta N}^{2}}{54\pi^{2}F_{\pi}^{2}}\{26+30 \ln(\frac{m_{\pi}}{\mu})+30 \frac{\Delta}{\sqrt{\Delta^2-m_{\pi}^{2}}}\ln[\frac{\Delta}{m_{\pi}}+
\sqrt{\frac{\Delta^{2}}{m_{\pi}^{2}}-1}]\}.
\end{array}
\end{equation}

\noindent
In the large $N_{c}$ limit, $\beta_{7}$ is simply $\frac{1}{6}$ \cite{TDC}
due to the same reason for simplification of $\beta_{17}$ \cite{KC}.

 On the other hand,
$b_{23}$ absorbs no divergence and its value is related to the 
axial mean square radius:
\begin{equation}
b_{23}=\frac{g_{A}}{6}\langle r^{2}_{A} \rangle .
\end{equation}

The data from (anti)neutrino-proton scattering
 gives $(4\pi F_{\pi})^{2}b_{23}$=3.08$\pm$0.27.
Besides the counterterms in ${\cal L}_{\pi N}$, the diagrams of t channel
 also involves the counterterms in ${\cal L}_{\pi\pi}$:
\begin{equation}
\begin{array}[c]{c}
 {\cal L}_{\pi\pi}^{(4)}=\frac{l_{3}}{16}Tr(\chi^{2}_{+})
+\frac{l_{4}}{16}\{2 Tr(D_{\mu}UD^{\mu}U^{\dagger}Tr(\chi_{-})^{2}+2
Tr(\chi^{\dagger}U\chi^{\dagger}U+\chi U^{\dagger}\chi U{\dagger})\\
-4Tr(\chi^{\dagger}\chi)-(Tr(\chi_{-})^{2}\}
+i\frac{l_{6}}{2}Tr([u^{\mu},u^{\nu}])f^{+}_{\mu\nu}+......
\end{array}
\end{equation}
 $l_{3}$ and $l_{4}$ appear in the chiral correction of pion
decay constant and pion mass respectively:
\begin{equation}
m_{\pi}^{2}=m_{0}^{2}+\frac{m_{0}^{4}}{F_{\pi}^{2}}(2l^{r}_{3}(\mu)+
\frac{1}{16\pi^{2}}\ln{\frac{m_{\pi}}{\mu}}).
\end{equation}
\begin{equation}
F_{\pi}=F_{0}+\frac{m_{0}^{2}}{F_{0}}(l_{4}^{r}(\mu)-\frac{1}{8\pi^{2}}
\ln{\frac{m_{\pi}}{\mu}}).
\end{equation}
$l_{6}$ only emerges at processes of pion electroproduction. It 
also absorbs the divergence:
\begin{equation}
{\it l}_{6}=-\frac{L}{6}+{\it l}^{r}_{6}(\mu).
\end{equation}
 Its finite part can be fixed by the empirical value of the pion mean
square charge radius:
\begin{equation}
\langle r^{2} \rangle_{\pi} =-\frac{1}{8\pi^{2}F_{\pi}^{2}}
(\ln\frac{m_{\pi}}{\mu}-\frac{12}{F_{\pi}^{2}}{\it l}_{6}^{r}(\mu)).
\end{equation}
Using the empirical values $\langle r^{2} \rangle_{\pi}$=0.439
$fm^2$ \cite{A}, we have ${\it l}_{6}(\mu=1 Gev)$=6.6 $\times 10^{-3}$.
Therefore the pion electroproduction at this order
 shares the same set of unknown parameters
and introduce none which cannot be independently fit by other processes.

\section {Result and Discussion}
The  results contain the s-channel $\Delta$(1232) pole.
As mentioned before the only known way to keep both unitarity and
the power counting scheme is to calculate
the weighted integrals as proposed
in \cite{KC}:
 
\begin{equation}
\bar{M}_{l\pm}^{(n)}=\frac{1}{m_{p}}\int_{m_{\pi}}^{E_{max}}
M_{l\pm}(E)(\frac{E}{m_{p}})^{n}dE.
\end{equation}

\noindent
The weighted integrals are parameterized as following:
\begin{equation}
\begin{array}[c]{c}
Re\bar{P}^{\pi^{0}P}_{i}=eg_{A}\zeta^{A}_{i}+eg_{A}\dot{\kappa}_{p}
\zeta^{B}_{i;B}+eg_{A}\kappa_{p}\zeta_{i;R}^{B}
+eg_{A}(1+\dot{\kappa}_{p})\tilde{c}_{1}\zeta^{C}_{i}\\ [.09in]
+eg_{A}^{3}\zeta^{D}_{i}+eg_{\pi \Delta N}\dot{G}_{1}\zeta^{E}_{i;B}
+eg_{\pi\Delta N}G_{1}\zeta_{i;R}^{E}
+eg_{\pi \Delta N}\tilde{G}_{2}\zeta^{F}_{i}+eg_{\pi\Delta N}\tilde{G}_{6}\zeta^{G}_{i}\\[.09in]
+e\tilde{g}_{\pi\Delta N}\dot{G}_{1}\zeta^{H}_{i}+eg_{\pi \Delta N}^{2}g_{A}\zeta^{K}_{i}+eg_{\pi \Delta N}^{2}g_{1}\zeta^{L}_{i}+e\tilde{b}_{9}\zeta^{M}_{i}
,\\[.09in]
i=1,2,3.
\end{array}
\end{equation}
\begin{equation}
\begin{array}[c]{c}
\frac{1}{\pi}Im\bar{P}^{\pi^{0}P}_{i}=eg_{A}^{3}\xi^{D}_{i}+eg_{\pi\Delta N}\dot{G}_{1}\xi^{E}_{i;B}+eg_{\pi\Delta N}G_{1}\xi^{E}_{i;R}+eg_{\pi\Delta N}\tilde{G}_{2}\xi^{F}_{i}+eg_{\pi\Delta N}\tilde{G}_{6}\xi^{G}_{i}\\[.09in]
+e\tilde{g}_{\pi\Delta N}\dot{G}_{1}\xi^{H}_{i}+eg_{\pi \Delta N}^{2}g_{A}\xi^{K}_{i}+eg_{\pi \Delta N}^{2}g_{1}\xi^{L}_{i},\\[.09in]
i=1,2,3.
\end{array}
\end{equation}

\noindent
Here $\tilde{G}_{2}$=$G_{2}+4G_{4}$, $\tilde{G}_{6}$=$G_{6}-G_{4}$, $\tilde{c}_{1}$=
$m_{N}c_{1}$, $\tilde{b}_{9}$=$b_{9}-b_{10}-
\frac{(4\pi F_{\pi})^{2})}{6m_{N}^{2}}g_{\pi\Delta N}
\dot{G}_{1}(1+4x+4z+12xz)$. $\dot{\kappa}_{p}$ means the 
parameter is taken in the limit:
$\Delta \rightarrow 0$, 
$m_{\pi}\rightarrow 0$, $\frac{\Delta}{m_{\pi}}$ fixed.
The first four terms in (18) are from tree graphs without the delta; the 
sixth to eleventh terms are due to tree graphs with the delta. Note that such  
tree graphs also contribute to the imaginary parts of amplitudes due 
to the delta function in $\frac{1}{E-\Delta+i\epsilon}$. The fifth 
term is from loop graphs without delta; the twelfth and thirteenth terms
are $\Delta-\pi$ loop contributions; the last term, which only 
appears in $P_{3}$, is due to the counterterms in ${\cal L}^{(3)}_{\pi NN}$.  
Note that the quantities,  
such as $\xi^{K}_{i},\xi^{L}_{i}$ are $\mu$-dependent, however final 
amplitudes are independent of $\mu$ because the $\kappa_{v}$, $G_{1}$,
 $\tilde{G}_{2}$ 
and $\tilde{G}_{6}$ are also $\mu$-dependent, and compensate the ones from the loop.\\

Similarly the longitude multipoles $L_{1+}$ and $L_{1-}$ also
 suffer from the same s--channel $\Delta$(1232) pole, therefore the same method is applied to
them and the weighted integrals are parameterized as:

\begin{equation}
\begin{array}[c]{c}
Re\bar{L}^{\pi^{0}P}_{1\pm}=eg_{A}\zeta^{A}_{\pm}+
eg_{A}(1+\dot{\kappa}_{p})\tilde{c}_{1}\zeta^{C}_{\pm}
+eg_{A}^{3}\zeta^{D}_{\pm}+eG_{\pi\Delta N}\dot{G}_{1}\zeta^{E}_{\pm;B}\\
+eg_{\pi \Delta N}\tilde{G}_{2}\zeta^{F}_{\pm}
+eg_{\pi \Delta N}^{2}g_{A}\zeta^{K}_{\pm}+eg_{\pi \Delta N}^{2}g_{1}\zeta^{L}_{\pm}.
\\
\end{array}
\end{equation}

\begin{equation}
\begin{array}[c]{c}
\frac{1}{\pi}Im\bar{L}^{\pi^{0}P}_{1\pm}=eg_{A}^{3}\xi^{D}_{\pm}+eg_{\pi\Delta N}\dot{G}_{1}\xi^{E}_{\pm;B}+eg_{\pi\Delta N}\tilde{G}_{2}\xi^{F}_{\pm}\\
+eg_{\pi \Delta N}^{2}g_{A}\xi^{H}_{\pm}+eg_{\pi \Delta N}^{2}g_{1}\xi^{K}_{\pm}.\\
\end{array}
\end{equation}

\noindent
Note that there is no ${\cal O}(\epsilon^{2})$ piece in $L_{1+}$, and the
${\cal O}(\epsilon^{2})$ piece of $L_{1-}$ is entirely due to the nucleon.
\\
  
Here we set $\Delta$=294 Mev and $F_{\pi}$=92.4 Mev, $M_{N}$=938.7 Mev, $E_{max}$
=340 Mev and $\mu$=500 Mev. All quantities are in the unit of $10^{-4}/m_{\pi}$
:
\newpage
\begin{center}
\begin{flushleft}
\vspace{2. mm}
\begin{small}
\begin{tabular}{|c|c|c|c|c|c|c|c|c|c|c|c|c|c|c|}
\hline
n&$\zeta^{A}_{1}$&$\zeta^{B}_{1;R}$&$\zeta^{R}_{1;B}$&$\zeta^{C}_{1}$&$\zeta_{1}^{D}$&$\zeta_{1;R}^{E}$&$\zeta_{1;B}^{E}$&$\zeta_{1}^{F}$&$\zeta_{1}^{G}
$&$\zeta_{1}^{H}$&$\zeta_{1;\pi}^{K}$&$\zeta_{1;\Delta}^{K}$&$\zeta_{1;\pi}^{L}$&$\zeta_{1;\Delta}^{L}$\\
\hline
1&24.80&29.26&-2.31&2.83&11.34&-12.97&-4.34&0.78&-3.13&-3.13
&-7.58&5.64&4.83&0\\
\hline
2&6.81&8.12&-0.71&0.78&3..34&-4.04&-1.14&0.31&-1.23&-1.23
&-1.64&1.53&2.03&0\\
\hline
3&1.95&2.35&-0.22&0.22&1.01&-1.76&-0.31&0.11&0.08&0.08&-0.36
&0.42&1.14&0\\
\hline
\end{tabular}
\end{small}
\noindent
\begin{small}
\begin{tabular}{|c|c|c|c|c|c|c|c|c|c|c|}
\hline
n&$\xi^{D}_{1}$&$\xi^{E}_{1;R}$&$\xi^{E}_{1;B}$&$\xi_{1}^{F}$&$\xi^{G}_{1}
$&$\xi^{H}_{1}$&$\xi^{K}_{1;\pi}$&$\xi_{1;\Delta}^{K}$&$\xi_{1;\pi}^{L}$&$\xi_{1;\Delta}^{L}$\\
\hline
1&2.51&6.64&-0.90&-0.45&1.80&1.80&-0.16&-2.19&0&-3.71\\
\hline
2&0.78&1.80&-0.25&-0.12&0.49&0.49&0.03&-0.60&0&-1.00\\
\hline
3&0.25&0.49&-0.07&-0.03&0.13&0.13&0.03&-0.16&0&-0.27\\
\hline
\end{tabular}
\end{small}
\noindent
\vspace{2. mm}
\begin{small}
\begin{tabular}{|c|c|c|c|c|c|c|c|c|c|c|c|c|c|c|}
\hline
n&$\zeta_{2}^{A}$&$\zeta_{2;R}^{B}$&$\zeta_{2:B}^{B}$&$\zeta_{2}^{C}$&$\zeta_{2}^{D}$&$\zeta_{2;R}^{E}$&$\zeta_{2;B}^{E}$&$\zeta_{2}^{F}$&$\zeta_{2}^{G}
$&$\zeta^{H}_{2}$&$\zeta^{K}_{2;\pi}$&$\zeta_{2;\Delta}^{K}$&$\zeta_{2;\pi}^{L}$&$\zeta_{2;\Delta}^{L}$\\
\hline
1&-14.96&-29.26&2.31&-2.83&-16.18&12.97&5.83&1.57&3.13&3.13
&10.16&22.22&-2.56&0\\
\hline
2&-4.12&-8.12&0.71&-0.78&-4.58&4.04&2.03&0.62&1.23&1.23
&3.09&6.03&-1.04&0\\
\hline
3&-1.18&-2.35&0.22&-0.22&-1.35&1.76&0.69&0.22&-0.08&-0.08
&0.96&1.64&-0.38&0\\
\hline
\end{tabular}
\end{small}
\noindent
\begin{small}
\begin{tabular}{|c|c|c|c|c|c|c|c|c|c|c|}
\hline
n&$\xi_{2}^{D}$&$\xi_{2;R}^{E}$&$\xi_{2;B}^{E}$&$\xi_{2}^{F}$&$\xi_{2}^{G}
$&$\xi^{H}_{2}$&$\xi^{K}_{2;\pi}$&$\xi_{2;\Delta}^{K}$&$\xi_{2;\pi}^{L}$&$\xi_{2;\Delta}^{L}$\\
\hline
1&6.65&-6.64&-1.80&-0.90&-1.80&-1.80&8.12&-1.92&0&1.78\\
\hline
2&-2.01&-1.80&-0.49&-0.24&-0.49&-0.49&2.77&-0.52&0&0.48\\
\hline
3&-0.62&-0.49&-0.14&-0.06&-0.13&-0.13&1.19&-0.14&0&0.13\\
\hline
\end{tabular}
\end{small} 
\noindent
\vspace{2. mm}
\begin{small}
\begin{tabular}{|c|c|c|c|c|c|c|c|c|c|c|c|c|c|c|}
\hline
n&$\zeta_{3}^{A}$&$\zeta_{3;R}^{B}$&$\zeta_{3;B}^{B}$&$\zeta_{3}^{C}$&$\zeta_{3;R}^{E}$&$\zeta_{3;B}^{E}$&$\zeta_{3}^{F}$&$\zeta_{3}^{G}
$&$\zeta^{H}_{3}$&$\zeta_{3;\pi}^{K}$&$\zeta_{3;\Delta}^{K}$&$\zeta_{3;\pi}^{L}$&$\zeta_{3;\Delta}^{L}$&$\zeta_{3}^{M}$\\
\hline
1&-0.86&4.58&-9.12&-9.49&13.20&-1.17&1.24&9.90&9.90
&-6.86&-16.58&12.92&0&20.64\\
\hline
2&-0.30&1.31&-2.67&-2.53&6.27&-0.19&0.44&3.53&3.53
&-2.32&-4.50&7.17&0&5.96\\
\hline
3&-0.10&0.39&-0.81&-0.71&2.46&-0.02&0.15&1.21&1.21
&-0.77&-1.22&2.96&0&1.78\\
\hline
\end{tabular}
\end{small}
\noindent
\begin{small}
\begin{tabular}{|c|c|c|c|c|c|c|c|c|c|}
\hline
n&$\xi_{3;R}^{E}$&$\xi_{3:B}^{E}$&$\xi_{3}^{F}$&$\xi_{3}^{G}
$&$\xi^{H}_{3}$&$\xi_{3;\pi}^{K}$&$\xi_{3;\Delta}^{K}$&$\xi_{3;\pi}^{L}$&$\xi_{3;\Delta}^{L}$\\
\hline
1&-13.29&-0.90&-0.45&-3.36&-3.36&-4.90&2.33&0&-18.59\\
\hline
2&-3.60&-0.24&-0.12&-0.91&-0.91&-1.71&0.63&0&-5.05\\
\hline
3&-0.98&-0.07&-0.03&-0.25&-0.25&-0.58&0.17&0&-1.37\\
\hline
\end{tabular}
\end{small}
\noindent
\vspace{2. mm}
\begin{small}
\begin{tabular}{|c|c|c|c|c|c|c|c|c|}
\hline
n&$\zeta_{+}^{A}$&$\zeta_{+}^{D}$&$\zeta_{+;B}^{E}$&$\zeta_{+}^{F}$&$\zeta^{K}_{+;\pi}$&$\zeta_{+;\Delta}^{K}$&$\zeta_{+;\pi}^{L}$&$\zeta_{+;\Delta}^{L}$\\
\hline
1&1.64&-0.81&0.25&0.39&1.54&4.64&0.45&0\\
\hline
2&0.45&-0.21&0.15&0.16&0.62&1.26&0.19&0\\
\hline
3&0.13&-0.06&0.06&0.06&0.22&0.34&0.39&0\\
\hline
\end{tabular}
\end{small}
\begin{small}
\begin{tabular}{|c|c|c|c|c|c|c|c|}
\hline
n&$\xi_{+}^{D}$&$\xi^{E}_{+;B}$&$\xi_{+}^{F}$&$\xi_{+;\pi}^{K}$&$\xi^{K}
_{+;\Delta}$&$\xi_{+;\pi}^{L}$&$\xi_{+;\Delta}^{L}$\\
\hline
1&-0.69&-0.45&-0.23&8.60&-1.03&0&-0.33\\
\hline
2&-0.21&-0.12&-0.06&2.92&-0.28&0&-0.09\\
\hline
3&-0.06&-0.06&-0.02&1.02&-0.08&0&-0.02\\
\hline
\end{tabular}
\end{small}
\noindent
\vspace{2. mm}
\begin{small}
\begin{tabular}{|c|c|c|c|c|c|c|c|c|c|}
\hline
n&$\zeta_{-}^{A}$&$\zeta_{-}^{C}$&$\zeta_{-}^{D}$&$\zeta_{-;B}^{E}$&$\zeta_{-}^{F}$&$\zeta^{K}_{-;\pi}$&$\zeta_{-;\Delta}^{K}$&$\zeta_{-;\pi}^{L}$&$\zeta_{-;\Delta}^{L}$\\
\hline
1&1.20&-8.44&-8.88&1.29&2.56&6.29&0&0.46&0\\
\hline
2&0.29&-2.25&-2.30&0.61&0.91&1.85&0&0.17&0\\
\hline
3&0.08&-0.63&-0.61&0.23&0.34&0.57&0&-0.59&0\\
\hline
\end{tabular}
\end{small}
\noindent
\begin{small}
\begin{tabular}{|c|c|c|c|c|c|c|c|c|c|}
\hline
n&$\xi_{-}^{D}$&$\xi_{-;B}^{E}$&$\xi_{-}^{F}$&$\xi^{K}_{-;\pi}$&$\xi_{-;\Delta}^{K}$&$\xi_{-;\pi}^{L}$&$\xi_{-;\Delta}^{L}$\\
\hline
1&-6.65&-1.36&-1.38&3.03&0.25&0&-0.28\\
\hline
2&-1.98&-0.36&-0.42&0.97&0.08&0&-0.08\\
\hline
3&-0.74&-0.11&-0.17&0.04&0.01&0&-0.03\\
\hline
\end{tabular}
\end{small}
\end{flushleft}
\end{center}

\vspace{5. mm}
\noindent {\sf Table 1: $Q^2$=0.01 $(Gev/c)^2$}
 
\newpage
\begin{center}
\begin{flushleft}
\vspace{2. mm}
\begin{small}
\begin{tabular}{|c|c|c|c|c|c|c|c|c|c|c|c|c|c|c|}
\hline
n&$\zeta^{A}_{1}$&$\zeta^{B}_{1;R}$&$\zeta^{R}_{1;B}$&$\zeta^{C}_{1}$&$\zeta_{
1}^{D}$&$\zeta_{1;R}^{E}$&$\zeta_{1;B}^{E}$&$\zeta_{1}^{F}$&$\zeta_{1}^{G}
$&$\zeta_{1}^{H}$&$\zeta_{1;\pi}^{K}$&$\zeta_{1;\Delta}^{K}$&$\zeta_{1;\pi}^{L}$&$\zeta_{1;\Delta}^{L}$\\
\hline
1&25.27&35.73&-7.96&4.21&13.82&-16.07&-5.57&0.93&-3.70&-3.70
&-9.31&2.11&7.65&0\\
\hline
2&6.87&9.79&-2.23&1.13&4.04&-5.89&-1.49&0.35&-1.39&-1.39
&-1.96&0.54&2.99&0\\
\hline
3&1.95&2.80&-0.65&0.32&1.22&-2.03&-0.42&0.12&0.10&0.10&-0.42
&0.14&1.06&0\\
\hline
\end{tabular}
\end{small}
\noindent
\begin{small}
\begin{tabular}{|c|c|c|c|c|c|c|c|c|c|c|}
\hline
n&$\xi^{D}_{1}$&$\xi^{E}_{1;R}$&$\xi^{E}_{1;B}$&$\xi_{1}^{F}$&$\xi^{G}_{1}
$&$\xi^{H}_{1}$&$\xi^{K}_{1;\pi}$&$\xi_{1;\Delta}^{K}$&$\xi_{1;\pi}^{L}$&$\xi_{1;\Delta}^{L}$\\
\hline
1&3.69&7.27&-0.94&-0.47&1.88&1.88&-0.74&-2.83&0&-4.74\\
\hline
2&1.11&1.88&-0.24&-0.12&0.49&0.49&-0.21&-0.73&0&-1.23\\
\hline
3&0.34&0.49&-0.06&-0.03&0.13&0.13&-0.06&-0.19&0&-0.32\\
\hline
\end{tabular}
\end{small}
\noindent
\vspace{2. mm}
\begin{small}
\begin{tabular}{|c|c|c|c|c|c|c|c|c|c|c|c|c|c|c|}
\hline
n&$\zeta_{2}^{A}$&$\zeta_{2;R}^{B}$&$\zeta_{2:B}^{B}$&$\zeta_{2}^{C}$&$\zeta_{2}^{D}$&$\zeta_{2;R}^{E}$&$\zeta_{2;B}^{E}$&$\zeta_{2}^{F}$&$\zeta_{2}^{G}
$&$\zeta^{H}_{2}$&$\zeta^{K}_{2;\pi}$&$\zeta_{2;\Delta}^{K}$&$\zeta_{2;\pi}^{L}$&$\zeta_{2;\Delta}^{L}$\\
\hline
1&-18.99&-35.73&7.96&-4.21&-19.37&16.07&6.91&1.87&3.70&3.70
&13.53&20.62&-4.95&0\\
\hline
2&-5.15&-9.79&2.23&-1.13&-5.44&5.89&2.34&0.70&1.39&1.39
&3.98&5.33&-1.88&0\\
\hline
3&-1.46&-2.80&0.65&-0.32&-1.59&2.03&0.77&0.24&-0.10&-0.10
&1.20&1.38&-0.66&0\\
\hline
\end{tabular}
\end{small}
\noindent
\begin{small}
\begin{tabular}{|c|c|c|c|c|c|c|c|c|c|c|}
\hline
n&$\xi_{2}^{D}$&$\xi_{2;R}^{E}$&$\xi_{2;B}^{E}$&$\xi_{2}^{F}$&$\xi_{2}^{G}
$&$\xi^{H}_{2}$&$\xi^{K}_{2;\pi}$&$\xi_{2;\Delta}^{K}$&$\xi_{2;\pi}^{L}$&$\xi_{2;\Delta}^{L}$\\
\hline
1&-8.51&-7.27&-1.88&-0.94&-1.88&-1.88&9.38&-2.12&0&2.74\\
\hline
2&-2.54&-1.88&-0.49&-0.24&-0.49&-0.49&3.08&-0.55&0&0.71\\
\hline
3&-0.78&-0.49&-0.12&-0.06&-0.13&-0.13&1.84&-0.14&0&0.18\\
\hline
\end{tabular}
\end{small} 
\noindent
\noindent
\vspace{2. mm}
\begin{small}
\begin{tabular}{|c|c|c|c|c|c|c|c|c|c|c|c|c|c|c|}
\hline
n&$\zeta_{3}^{A}$&$\zeta_{3;R}^{B}$&$\zeta_{3;B}^{B}$&$\zeta_{3}^{C}$&$\zeta_{3;R}^{E}$&$\zeta_{3;B}^{E}$&$\zeta_{3}^{F}$&$\zeta_{3}^{G}
$&$\zeta^{H}_{3}$&$\zeta_{3;\pi}^{K}$&$\zeta_{3;\Delta}^{K}$&$\zeta_{3;\pi}^{L}$&$\zeta_{3;\Delta}^{L}$&$\zeta_{3}^{M}$\\
\hline
1&-3.59&7.78&-15.60&-10.18&16.61&-1.18&1.47&11.79&11.79
&-9.21&-18.48&17.94&0&24.09\\
\hline
2&-1.06&2.14&-4.41&-2.71&7.41&-0.19&0.51&4.06&4.06
&-2.88&-4.50&8.99&0&6.95\\
\hline
3&-0.32&0.62&-1.30&-0.76&2.78&-0.02&0.17&1.35&1.35
&-0.90&-1.24&3.51&0&2.06\\
\hline
\end{tabular}
\end{small}
\noindent
\begin{small}
\begin{tabular}{|c|c|c|c|c|c|c|c|c|c|}
\hline
n&$\xi_{3;R}^{E}$&$\xi_{3:B}^{E}$&$\xi_{3}^{F}$&$\xi_{3}^{G}$
&$\xi^{H}_{3}$&$\xi_{3;\pi}^{K}$&$\xi_{3;\Delta}^{K}$&$\xi_{3;\pi}^{L}$&$\xi_{3;\Delta}^{L}$\\
\hline
1&-14.54&-0.94&-0.47&-2.90&-2.90&-5.73&2.09&0&-21.13\\
\hline
2&-3.76&-0.24&-0.12&-0.75&-0.75&-1.97&0.54&0&-5.46\\
\hline
3&-0.97&-0.06&-0.03&-0.19&-0.19&-0.65&0.14&0&-1.41\\
\hline
\end{tabular}
\end{small}
\noindent
\vspace{2. mm}
\begin{small}
\begin{tabular}{|c|c|c|c|c|c|c|c|c|}
\hline
n&$\zeta_{+}^{A}$&$\zeta_{+}^{D}$&$\zeta_{+;B}^{E}$&$\zeta_{+}^{F}$&$\zeta^{K}_{+;\pi}$&$\zeta_{+;\Delta}^{K}$&$\zeta_{+;\pi}^{L}$&$\zeta_{+;\Delta}^{L}$\\
\hline
1&1.04&-0.93&0.22&0.47&2.51&3.79&0.54&0\\
\hline
2&0.29&-0.23&0.15&0.18&0.91&0.98&0.22&0\\
\hline
3&-0.08&-0.06&0.06&0.06&0.31&0.25&0.08&0\\
\hline
\end{tabular}
\end{small}
\noindent
\begin{small}
\begin{tabular}{|c|c|c|c|c|c|c|c|}
\hline
n&$\xi_{+}^{D}$&$\xi^{E}_{+;B}$&$\xi_{+}^{F}$&$\xi^{K}_{+;\pi}$
&$\xi_{+;\Delta}^{K}$&$\xi_{+;\pi}^{L}$&$\xi_{+;\Delta}^{L}$\\
\hline
1&-0.80&-0.47&-0.24&6.62&-1.24&0&-0.35\\
\hline
2&-0.24&-0.12&-0.06&2.05&-0.32&0&-0.09\\
\hline
3&-0.19&-0.04&-0.02&0.85&-0.08&0&-0.02\\
\hline
\end{tabular}
\end{small}
\noindent
\vspace{2. mm}
\begin{small}
\begin{tabular}{|c|c|c|c|c|c|c|c|c|c|}
\hline
n&$\zeta_{-}^{A}$&$\zeta_{-}^{C}$&$\zeta_{-}^{D}$&$\zeta_{-;B}^{E}$&$\zeta_{-}^{F}$&$\zeta^{K}_{-;\pi}$&$\zeta_{-;\Delta}^{K}$&$\zeta_{-;\pi}^{L}$&$\zeta_{-;\Delta}^{L}$\\
\hline
1&-6.00&-11.95&-10.18&1.39&2.77&7.81&0&-3.45&0\\
\hline
2&-1.69&-3.12&-2.58&0.63&1.04&2.29&0&-1.00&0\\
\hline
3&-0.50&-0.86&-0.67&0.25&0.37&0.69&0&-0.30&0\\
\hline
\end{tabular}
\end{small}
\noindent
\begin{small}
\begin{tabular}{|c|c|c|c|c|c|c|c|}
\hline
n&$\xi_{-}^{D}$&$\xi_{-;B}^{E}$&$\xi_{-}^{F}$&$\xi^{K}_{-;\pi}$&$\xi_{-;\Delta}^{K}$&$\xi_{-;\pi}^{L}$&$\xi_{-;\Delta}^{L}$\\
\hline
1&-7.49&-1.41&-1.41&2.09&0.29&0&-0.31\\
\hline
2&-1.85&-0.36&-0.36&0.65&0.07&0&-0.07\\
\hline
3&-0.69&-0.11&-0.10&0.47&0.02&0&-0.01\\
\hline
\end{tabular}
\end{small}
\end{flushleft}
\end{center}

\vspace{5. mm}
\noindent{\sf Table 2: $Q^2$=0.04 $(Gev/c)^2$} 
\newpage
\begin{center}
\begin{flushleft}
\vspace{2. mm}
\begin{small}
\begin{tabular}{|c|c|c|c|c|c|c|c|c|c|c|c|c|c|c|}
\hline
n&$\zeta^{A}_{1}$&$\zeta^{B}_{1;R}$&$\zeta^{R}_{1;B}$&$\zeta^{C}_{1}$&$\zeta_{
1}^{D}$&$\zeta_{1;R}^{E}$&$\zeta_{1;B}^{E}$&$\zeta_{1}^{F}$&$\zeta_{1}^{G}
$&$\zeta_{1}^{H}$&$\zeta_{1;\pi}^{K}$&$\zeta_{1;\Delta}^{K}$&$\zeta_{1;\pi}^{L}$&$\zeta_{1;\Delta}^{L}$\\
\hline
1&23.67&39.89&-13.59&5.27&15.63&-18.35&-6.41&1.02&-4.10&-4.10
&-10.14&2.17&9.98&0\\
\hline
2&6.44&10.86&-3.70&1.39&4.56&-6.52&-1.73&0.37&-1.49&-1.49
&-2.11&0.54&3.71&0\\
\hline
3&1.83&3.09&-1.06&0.11&1.37&-2.19&-0.49&0.13&0.11&0.11&-0.46
&0.14&1.27&0\\
\hline
\end{tabular}
\end{small}
\noindent
\begin{small}
\begin{tabular}{|c|c|c|c|c|c|c|c|c|c|c|}
\hline
n&$\xi^{D}_{1}$&$\xi^{E}_{1;R}$&$\xi^{E}_{1;B}$&$\xi_{1}^{F}$&$\xi^{G}_{1}
$&$\xi^{H}_{1}$&$\xi^{K}_{1;\pi}$&$\xi_{1;\Delta}^{K}$&$\xi_{1;\pi}^{L}$&$\xi_{1;\Delta}^{L}$\\
\hline
1&4.01&7.55&-0.94&-0.47&1.88&1.88&-0.71&-3.22&0&-5.38\\
\hline
2&1.20&1.89&-0.24&-0.12&0.47&0.47&-0.21&-0.80&0&-1.35\\
\hline
3&0.37&0.47&-0.06&-0.03&0.12&0.12&-0.06&-0.20&0&-0.01\\
\hline
\end{tabular}
\end{small}
\noindent
\vspace{2. mm}
\begin{small}
\begin{tabular}{|c|c|c|c|c|c|c|c|c|c|c|c|c|c|c|}
\hline
n&$\zeta_{2}^{A}$&$\zeta_{2;R}^{B}$&$\zeta_{2:B}^{B}$
&$\zeta_{2}^{C}$&$\zeta_{2}^{D}$&$\zeta_{2;R}^{E}$&$\zeta_{2;B}^{E}$&$\zeta_{2}^{F}$&$\zeta_{2}^{G}
$&$\zeta^{H}_{2}$&$\zeta^{K}_{2;\pi}$&$\zeta_{2;\Delta}^{K}$&$\zeta_{2;\pi}^{L}$&$\zeta_{2;\Delta}^{L}$\\
\hline
1&-21.72&-39.89&13.59&-5.27&-21.44&18.35&7.62&2.07&4.10&4.10
&15.84&20.53&-6.96&0\\
\hline
2&-5.84&-10.86&3.70&-1.39&-6.00&6.52&2.53&0.75&1.49&1.49
&4.56&5.13&-2.54&0\\
\hline
3&-1.64&-3.09&1.06&-0.11&-1.75&2.19&0.82&0.26&-0.11&-0.11
&1.35&1.28&-0.86&0\\
\hline
\end{tabular}
\end{small}
\noindent
\begin{small}
\begin{tabular}{|c|c|c|c|c|c|c|c|c|c|c|}
\hline
n&$\xi_{2}^{D}$&$\xi_{2;R}^{E}$&$\xi_{2;B}^{E}$&$\xi_{2}^{F}$&$\xi_{2}^{G}
$&$\xi^{H}_{2}$&$\xi^{K}_{2;\pi}$&$\xi_{2;\Delta}^{K}$&$\xi_{2;\pi}^{L}$&$\xi_{2;\Delta}^{L}$\\
\hline
1&-9.84&-7.55&-1.89&-0.95&-1.88&-1.88&10.33&-2.14&0&3.37\\
\hline
2&-2.92&-1.89&-0.49&-0.24&-0.49&-0.49&3.33&-0.54&0&0.84\\
\hline
3&-0.89&-0.47&-0.12&-0.06&-0.13&-0.13&1.35&-0.13&0&0.21\\
\hline
\end{tabular}
\end{small} 
\noindent
\vspace{2. mm}
\begin{small}
\begin{tabular}{|c|c|c|c|c|c|c|c|c|c|c|c|c|c|c|}
\hline
n&$\zeta_{3}^{A}$&$\zeta_{3;R}^{B}$&$\zeta_{3;B}^{B}$&$\zeta_{3}^{C}$&$\zeta_{3;R}^{E}$&$\zeta_{3;B}^{E}$&$\zeta_{3}^{F}$&$\zeta_{3}^{G}
$&$\zeta^{H}_{3}$&$\zeta_{3;\pi}^{K}$&$\zeta_{3;\Delta}^{K}$&$\zeta_{3;\pi}^{L}$&$\zeta_{3;\Delta}^{L}$&$\zeta_{3}^{M}$\\
\hline
1&-6.77&10.32&-21.65&-10.86&19.41&-1.09&1.63&13.04&13.04
&-10.76&-18.35&22.36&0&26.47\\
\hline
2&-1.90&2.80&-6.00&-2.86&8.20&-0.17&0.55&4.38&4.38
&-3.21&-4.59&10.33&0&7.54\\
\hline
3&-0.55&0.80&-1.73&-0.23&2.97&-0.02&0.18&1.43&1.43
&-0.97&-1.15&3.87&0&2.23\\
\hline
\end{tabular}
\end{small}
\noindent
\begin{small}
\begin{tabular}{|c|c|c|c|c|c|c|c|c|c|}
\hline
n&$\xi_{3;R}^{E}$&$\xi_{3:B}^{E}$&$\xi_{3}^{F}$&$\xi_{3}^{G}$
&$\xi^{H}_{3}$&$\xi_{3;\pi}^{K}$&$\xi_{3;\Delta}^{K}$&$\xi_{3;\pi}^{L}$&$\xi_{3;\Delta}^{L}$\\
\hline
1&-15.11&-0.94&-0.47&-2.61&-2.61&-6.18&1.81&0&-22.47\\
\hline
2&-3.78&-0.24&-0.12&-0.65&-0.65&-2.08&0.45&0&-5.61\\
\hline
3&-0.94&-0.06&-0.03&-0.12&-0.12&-0.68&0.11&0&-1.40\\
\hline
\end{tabular}
\end{small}
\noindent
\vspace{2. mm}
\begin{small}
\begin{tabular}{|c|c|c|c|c|c|c|c|c|}
\hline
n&$\zeta_{+}^{A}$&$\zeta_{+}^{D}$&$\zeta_{+;B}^{E}$&$\zeta_{+}^{F}$&$\zeta^{K}_{+;\pi}$&$\zeta_{+;\Delta}^{K}$&$\zeta_{+;\pi}^{L}$&$\zeta_{+;\Delta}^{L}$\\
\hline
1&0.33&-0.97&0.20&0.52&3.22&6.71&0.61&0\\
\hline
2&0.10&-0.24&0.13&0.19&1.12&3.87&0.23&0\\
\hline
3&0.03&-0.06&0.06&0.06&0.37&0.42&0.08&0\\
\hline
\end{tabular}
\end{small}
\noindent
\begin{small}
\begin{tabular}{|c|c|c|c|c|c|c|c|}
\hline
n&$\xi_{+}^{D}$&$\xi_{+;B}^{E}$&$\xi_{+}^{F}$&$\xi^{K}_{+;\pi}$&$\xi_{+;\Delta}^{K}$&$\xi_{+;\pi}^{L}$&$\xi_{+;\Delta}^{L}$\\
\hline
1&-0.98&-0.47&-0.47&-1.31&6.70&0&-0.36\\
\hline
2&-0.29&-0.12&-0.12&-0.33&2.16&0&-0.09\\
\hline
3&-0.09&-0.03&-0.03&-0.08&0.74&0&-0.02\\
\hline
\end{tabular}
\end{small}
\noindent
\vspace{2. mm}
\begin{small}
\begin{tabular}{|c|c|c|c|c|c|c|c|c|c|}
\hline
n&$\zeta_{-}^{A}$&$\zeta_{-}^{C}$&$\zeta_{-}^{D}$&$\zeta_{-;B}^{E}$&$\zeta_{-}^{F}$&$\zeta^{K}_{-;\pi}$&$\zeta_{-;\Delta}^{K}$&$\zeta_{-;\pi}^{L}$&$\zeta_{-;\Delta}^{L}$\\
\hline
1&-10.49&-14.75&-10.65&1.43&3.08&8.87&0&-3.82&0\\
\hline
2&-2.83&-3.80&-2.65&0.64&1.12&2.62&0&-1.11&0\\
\hline
3&-0.80&-1.03&-0.69&0.23&0.38&0.80&0&-0.33&0\\
\hline
\end{tabular}
\end{small}
\noindent
\begin{small}
\begin{tabular}{|c|c|c|c|c|c|c|c|}
\hline
n&$\xi_{-}^{D}$&$\xi_{-;B}^{E}$&$\xi_{-}^{F}$&$\xi^{K}_{-;\pi}$&$\xi_{-;\Delta}^{K}$&$\xi_{-;\pi}^{L}$&$\xi_{-;\Delta}^{L}$\\
\hline
1&-8.58&-1.42&-1.42&2.54&0.27&0&-0.30\\
\hline
2&-2.55&-0.35&-0.35&0.78&0.07&0&-0.08\\
\hline
3&-0.78&-0.08&-0.08&0.34&0.02&0&-0.02\\
\hline
\end{tabular}
\end{small}
\end{flushleft}
\end{center}

\vspace{5. mm}
\noindent{\sf Table 3: $Q^2$=0.06 $(Gev/c)^2$}
\newpage
\noindent

Although we cannot pin down the values of unknown parameters such as
$G_{1}$, $G_{2}$...
since so far no amplitudes directly extracted from experiment are available,
 we still can make some observation on these integrated quantities 
which may shed the lights to the physics behind them.\\

 Firstly, the values of $\zeta^{A}_{i}$,$\zeta^{B}_{i;R}$ and $\zeta^{E}_{i;R}$
 are significant lager among the ones of tree diagrams because they
 associate with the ${\cal O}(\epsilon^2)$ amplitudes and others are related to
  ${\cal O}(\epsilon^3)$ amplitudes
 (except $\zeta^{B}_{3;R}$ because they are identically zero if
 recoil effect is not included).
 Therefore power counting scheme is well preserved here and
 the wild behavior of amplitudes due to the pole
 of $\Delta$(1232) resonance is tamed
 in the weighted integrals.
 The $Q^2$ dependences of these quantities of tree diagrams
 are less sensitive than ones of the loop diagrams, which is due to 
 up to ${\cal O}(\epsilon^{3})$ where there is no vertex proportional
 to $Q^2$. The $Q^2$ dependences of these multipole results of
tree diagrams are only from the propagators of the nucleon and $\Delta$(1232). 
 However if we continue to go to higher $Q^2$ range, power counting could not
be kept without modification. We made this approximation on the propagators
of the nucleon:

$$S(k)=\frac{i}{v\cdot k+\frac{(v\cdot k)^{2}-k^{2}}{2m_{p}}}
\sim \frac{i}{v\cdot k}+\frac{i((v\cdot k)^2-k^2)}{2m_{p}(v\cdot k)^{2}}.$$

\noindent
k is the four momentum of the nucleon.
Here we assume that $Q^2 \sim \epsilon^{2}$. 
When $Q^2$ goes higher, ${\it e.g}$,
 $Q^2 \sim \epsilon\Lambda \sim 0.3(Gev/c)^2$
, obviously this expansion will
break down or at least the counting rules have to be changed. 
 One way to do this is to calculate them in full relativistic
formulation then expand it by $\frac{\omega}{M_{p}}$ but not by $Q^2$.
So we have to limit ourselves in the region
 $Q^2 \leq 0.1 \sim 0.2(Gev/c)^2$. \\

  In general the weighted integrals
related to the loop diagrams, such as $\zeta^{D}_{i}$ ,$\zeta^{K}_{i}$
 and $\zeta^{L}_{i}$, are all more sensitive to the change of $Q^{2}$.
Their $Q^2$ dependences are from the propagators of pions in the N$\pi$
and $\Delta \pi$ loops.
Naively we might conclude that the effects of the
pion cloud are quite significant in the $Q^2$
evolution of the weighted integrals. However these quantities
are $\mu$-depedent and their $\mu$ dependences are
 compensated by other quantities. Therefore it is difficult to identify
the generic $\pi$ cloud effects. 
 On the other hand, the quantities such as $\zeta^{K}_{i;\Delta}$
which is free of any $\mu$ dependence are less sensitive to
$Q^2$. Such a quantity represents the interference between the imaginary 
part of the N$\pi$ loop and the $\delta$ function of the $\Delta$ propagator.
Again the validity of our expansion is limited in the low $Q^2$ region
because of the possible modification in
 the propagators of nucleon and $\Delta$(1232) in the loops.   
\\

 The longitudinal multipoles
$L_{1+}$ and $L_{1-}$ are now the subjects of intense study
 \cite{W,ST}.
 It was suggested that in parallel
 kinematics, the particular combination
$4S_{1+}+S_{1-}-S_{0+}$ can be measured through 
the recoil polarizations of the nucleon and attempts have been made to
extract the CMR=$ImS_{1+}/ImM_{1+}$ at the $\Delta$(1232) peak under the
assumption that the nonresonant contribution is negligible.
However, a recent measurement on $P_{n}$ which should vanish if the
background is really negligible at the peak, was reported
as unexpectedly large \cite{M}. The MIT/Bates group also measured
the longitudinal-transverse interference response $R_{LT}$ and the related
asymmetry $A_{LT}$ at $W$=1.172 Gev, and show that available models all fail to
explain the $W$-dependence of these observables.
\\

Our approach gives no information on $W$ dependence since we integrate through
the $\Delta$(1232) region. If more data is taken at a different energy
in the $\Delta$ region, then our result can be verified
 because no new counterterms are needed for both $L_{1+}$
and $L_{1-}$ up to ${\cal O}(\epsilon^{3})$.
 $L_{1+}$ is dominated by the $\Delta$
$\pi$ loop diagrams; on the other hand, the $L_{1-}$ is dominated by
the nucleon sector.
  There is one special interesting combination
 to note: $L_{1+}-E_{1+}$. It is 
exclusively due to the $\Delta \pi$ loop; further, it is a $\mu$-independent
quantity. It is interesting to see
if any observable is particularly sensitive to such a combination \cite{CK}. \\

  A complete
 determination of counterterms in the $\gamma \pi N\Delta$ system
can only be done by a more extensive study on more processes such as 
$\pi$N scattering or Compton scattering. For example, the contributions
of $g_{\pi\Delta N}\tilde{G}_{6}$ cannot be disentangled with the ones
of $\tilde{g}_{\pi\Delta N}\dot{G}_{1}$ in either photo- or electroproduction
of pions. But only in $\tilde{g}_{\pi\Delta N}$ participates the $\pi$N
scattering. Similarly only $\dot{G}_{1}$ and $\tilde{G}_{6}$ appear
in the Compton scattering. Actually they share most of the counterterms
and that is why even we could make no prediction on any physical observables.
 Still our approach has predictive power.
 A similar calculation on $\pi$N
elastic scattering is proceeding \cite{CK}.
\\

Finally, we emphasized that
 in order to
 compare with our results, we require these double
polarization experiments continuously run for a  wide range of energy
and eventually to extract the individual amplitudes.
 To describe both longitudinal and transverse amplitudes of
pion electroproduction in the resonance region
including the resonant and background contributions are a formidable
task. Our approach seem to be a very promising way, at least in the low $Q^2$
region.\\

\section {Conclusion}
 In summary, we calculated the amplitudes of pion electroproduction
up to third order in the framework of HBChPT including explicit $\Delta$
(1232) isobars.
 The $Q^2$ dependences of the weighted integrals of these amplitudes
are presented here and
 they turn out to be good testing grounds for this phenomenological
extension of HBChPT because all the counterterms are fixed at $Q^2$=0.
 The predictive power
of the method proposed in \cite{KC} was shown.
 Our final goal is to treat various processes such
as $\gamma N\rightarrow \gamma N$, $\gamma N\rightarrow \gamma N$ and
$\pi N\rightarrow \pi N$ in the $\Delta$(1232) region 
in a unified, consistent framework, and determine all counterterms
from experiments without any model dependence,
 which requires
 calculations on more processes and the results of
 multipoles analyses extracted from experimental 
data. \\

\section {Acknowledgment}
The support of the U.S Department of Energy under grant DE-FG02-93ER-40762
 is gratefully
acknowledged.

\end{document}